# 11 CMOS-Compatible Nanowire Biosensors

*Thanh C. Nguyen, Wanzhi Qiu, Matteo Altissimo, Paul G. Spizzirri, Laurens H. Willems van Beveren, and Efstratios Skafidas*

## CONTENTS

## 11.1 INTRODUCTION

Seralogical point-of-care diagnostic tests that are able to detect disease-causing viruses, bacteria, antigens, or prions are required for clinical diagnosis. Beyond blood serum, the development of robust and inexpensive protocols that can operate in widely varying environments (e.g., biological and physical) can prove even more challenging. Underpinning many conventional detection protocols that include enzyme-linked immunosorbent assay (ELISA) techniques is the use of biological antibodies due to their unique selectivity to target antigens. Being able to directly determine the binding status of these efficient molecular detectors through some form of electronic readout has become an active research area for lab-on-a-chip (LOC) applications.

Silicon nanowire devices, which are effectively low- (or one-) dimensional resistive semiconductor channels, display conductance sensitivity to surface charges (and their resulting electric fields) due to their high surface area-to-volume ratio. When an external charge is in close proximity to the surface of the nanowire, it modifies the charge carrier distribution in the channel, which in turn substantially affects the electrical properties: conductance, quantum capacitance, and kinetic inductance. Since it is also known that biological molecules can exhibit a net charge due to ionized groups at pH values far from their isoelectric point, biomolecules attached to the surface of nanowires should be capable of influencing the electronic response of such a device.

Nanowire biosensors designed using this principle, namely, measuring changes in conductance, have been reported recently [1–12]. Gengfeng et al. recently reported a direct, real-time electrical detection of a single influenza A virus [12], while Hahm, Lieber, and Gao have demonstrated the use of a PNA probe to detect a particular DNA sequence with a limit of detection of 10 fM [1,9]. The methods reported in the literature are based upon direct measurement of the change in DC conductance of the nanowire due to the attachment of a molecule of interest to the functionalized surface. Unfortunately, measurements of small changes in conductance can be difficult and require sensitive, low-noise amplifiers and high-resolution, analog-to-digital converters. Very low-noise, low-frequency, and high-gain amplifiers are difficult to implement on small-geometry complementary metal-oxide-semiconductor (CMOS) processes because of the inherently high value of the low-frequency flicker noise [13].

Recent results on nanowire detection are presented here that indicate that silicon nanowires exhibit a frequency-dependent transfer function that resembles that of a high-pass filter. In order to better understand this response, we describe in this chapter methods to model and simulate the frequency response of a three-dimensional silicon nanowire (SiNW) field effect transistor (FET) biosensor. We will show using these models that as biomolecules with a higher net charge attach to the nanowire, they displace more charge carriers within the nanowire channel, causing the corner frequency (i.e., the location of the 3 dB point of the transfer function) of the filter to decrease along with the conductance, quantum capacitance, and kinetic inductance. This property of silicon nanowires, which was first shown in [14], is further developed in this work to build a low-cost CMOS, frequency-based detection system.

In addition to simulating the device response, we also consider factors that affect the high-pass filter behavior of these nanostructures in addition to methods that could be used to help quantify the amount of the antigen present in the sample. Competitive versus captive antigen binding scenarios can also change the analysis paradigm from that of single positive detection to quantitative analysis, depending upon the clinical need. Simulations in this work have shown, however, that competitive binding, which results in a high number of antibody/antigen attachments per unit time, gives rise to the frequency response of the device, and that the location of the corner frequency of the high-pass filter varies with the average number of these attachments. Therefore practical realization of such a sensor requires the development of readout circuitry capable of processing the SiNW FET frequency-dependent characteristic in real time. Lastly, several CMOS circuit designs that could be used to determine the amount of charge attached to the nanowire are presented and discussed.

## 11.2   DEVICE CONSTRUCTION

Although many fabrication techniques exist for fabricating silicon nanowires, most are not compatible with planar CMOS fabrication processes. The top-down method [1], however, which results in the device being fabricated on a thin device layer atop a silicon-on-insulator (SOI) wafer, is compatible with CMOS fabrication processes.

The silicon dioxide insulator layer (buried oxide) is approximately 150 nm thick and provides both electrical isolation and a reduction in parasitic device capacitance. The top silicon device layer, which has a thickness of 50 nm, is phosphorus doped to a concentration of 1.0 $e^{15}$ cm$^{-3}$ to produce a semiconducting nanowire using low-energy (<30 keV) ion implantation techniques. Photoresist, negative tone electron-beam resist, and electron-beam lithography are used to fashion the nanowire (NW) structures on the top of the SOI, and the isotropic reactive ion etching technique is finally used to remove the nonmasked areas, leaving a fin structure as shown in Figure 11.1. The two contact ends of the nanowire must be further doped to a higher concentration of phosphorus, creating local n$^+$ regions capable of producing near-ohmic contact to the metallic bonding pad on top. The next step involves performing a rapid thermal anneal (950°C for 5 s) to activate the phosphorus dopants. Finally, metallic contact pads are deposited at the two ends of the nanowire to form the ohmic source and drain contacts, and these are spike annealed in forming gas. Antibody conjugation to the device surface is performed using procedures described in later sections in order to produce the chemical gate. Figure 11.2 illustrates the final nanowire design along with the simulated doping profiles.

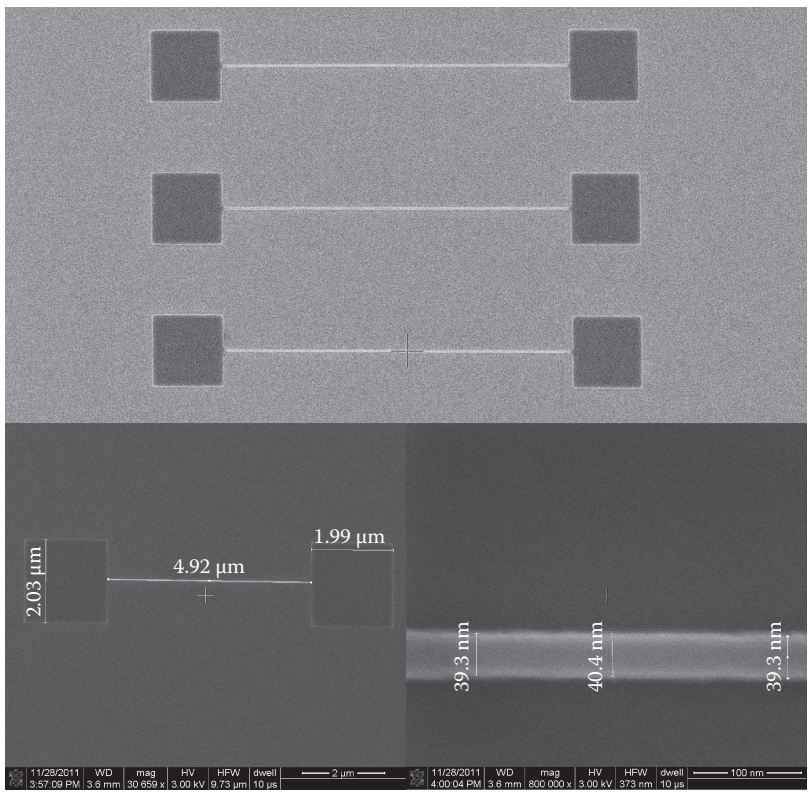

**FIGURE 11.1** Silicon-on-insulator (SOI)-based nanowires (40 nm in width and ~5 um long) fabricated using e-beam lithography followed by RIE etching.

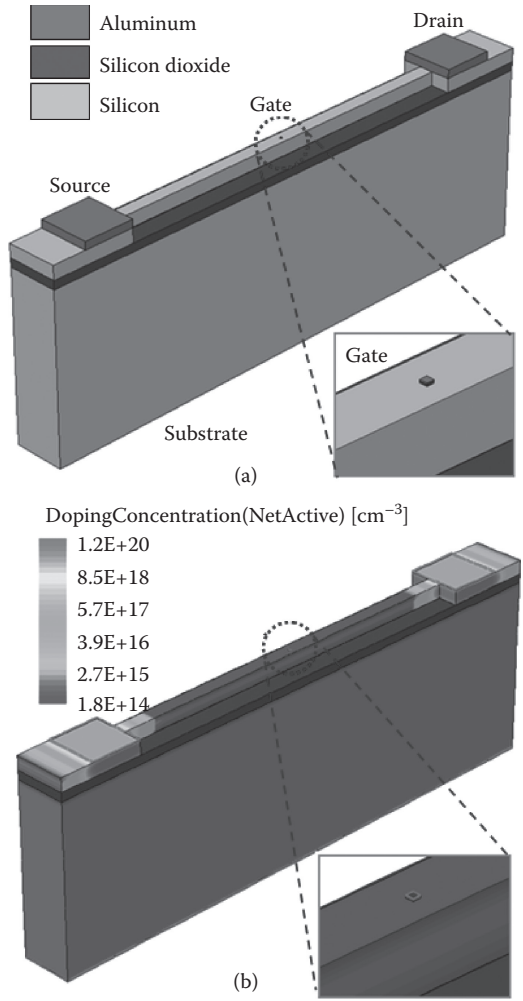

**FIGURE 11.2** (a) SiNW FET device structure material view. (The insets show a magnified view of the gate on the channel surface.) (b) Doping concentration view.

## 11.3 DC CHARACTERISTICS OF AN N-TYPE SINW FET

The sensitivity of the silicon nanowire is defined to be the ratio between the change and initial conductance $S = \Delta G/G_0$ [2]. When the drain-to-source voltage is constant, the sensitivity can be written as $S = |I - I_0|/I_0$, where $I_0$ is the current at zero charge and $|I - I_0|$ is the current change upon the attachment of charge to the nanowire surface. Figure 11.3(c) shows the sensitivity versus substrate voltage with surface charge varied from –5 $e$ to +5 $e$, where $e$ is the fundamental unit of charge. It can be seen from Figure 11.3(b) and (c) that negative charges on the surface of an n-type silicon nanowire exhibit greater sensitivity than when positive charges are attached to the same surface. This result is in agreement with the nanowire field effect transistor

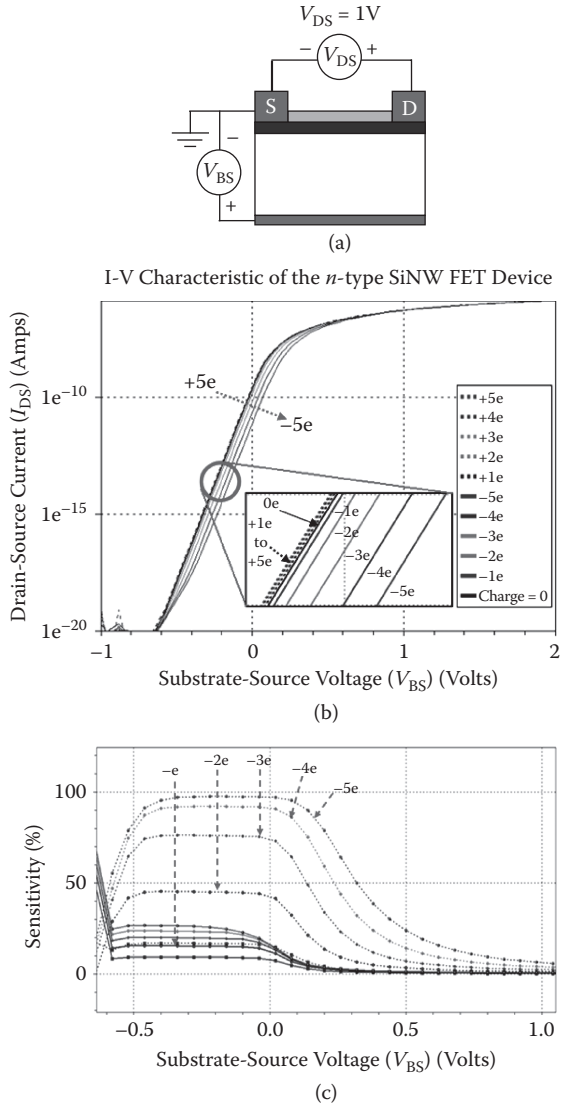

**FIGURE 11.3** (a) SiNW FET device DC schematic. (b) I-V characteristics as a function of gate net charge. Substrate-source voltage is swept from –1 V to 2 V while drain-source bias is 1 V. (c) Device sensitivity as a function of gate charge.

MATLAB®* simulation of Nair and Alam [18], whose simulations were performed in air. In the same work, it was suggested that the device sensitivity should be reduced when the NW is in aqueous solution because of the high dielectric constant of water, which will reduce the depletion depth into the NW body [18].

---

* MATLAB® is a registered trademark of The MathWorks, Inc.

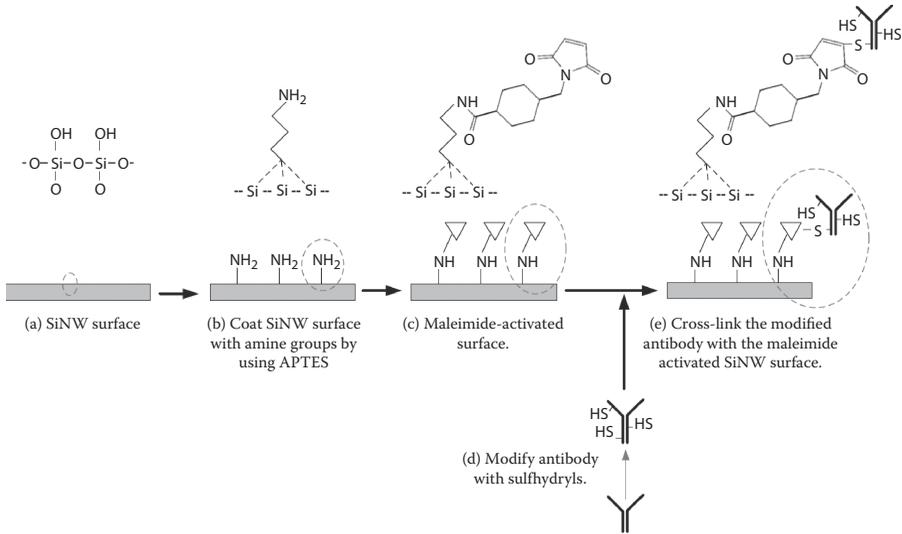

**FIGURE 11.4** Steps involved in nanowire functionalization. Processing of the nanowire surface in order to conjugate antibodies that function as chemical gates.

## 11.4 NANOWIRE FUNCTIONALIZATION PROTOCOL

Protocols for functionalizing the nanowire can be found in [1,2,4–14,16,17]. In this work, the protocol for functionalization is shown in Figure 11.4 and is based on the method described in [15]. The functionalization of the nanowire surface is achieved by first thoroughly washing the device with a mixture of acetone and ethanol (1:1 v/v). After drying, the nanowire is silylated using an aminosilane reagent (3-amino-propyltriethoxysilane (APTES)). This step results in a nanowire surface that is coated with amine groups. A cross-linker (Sulfo-SMCC) is then applied to the silylated surface and activated with maleimide. Sulfhydryl groups are made available on an antibody by using Traut's reagent, which reacts with primary amines ($-NH_2$) present on the side chain of lysine residues of antibodies, creating sulfhydryl groups upon reaction with the maleimide-activated surface. The next step is to immediately cover the maleimide-activated surface with the modified antibody solution, which is then incubated for 2–4 h at room temperature for the antibody to covalently attach to the nanowire surface. The surface is then thoroughly rinsed with coupling buffer phosphate-buffered saline (PBS). After this step, the nanowire and antibody system is ready for use as a detection assay.

## 11.5 FREQUENCY-DEPENDENT METHOD FOR BIOMOLECULE DETECTION

The DC change in conduction has been described for biomolecule sensing purposes [1]. Although the relative change in conduction can be large, the absolute conductance is small, requiring large amplification of the signal received across the nanowire. The resulting signal can be noisy and difficult to determine accurately.

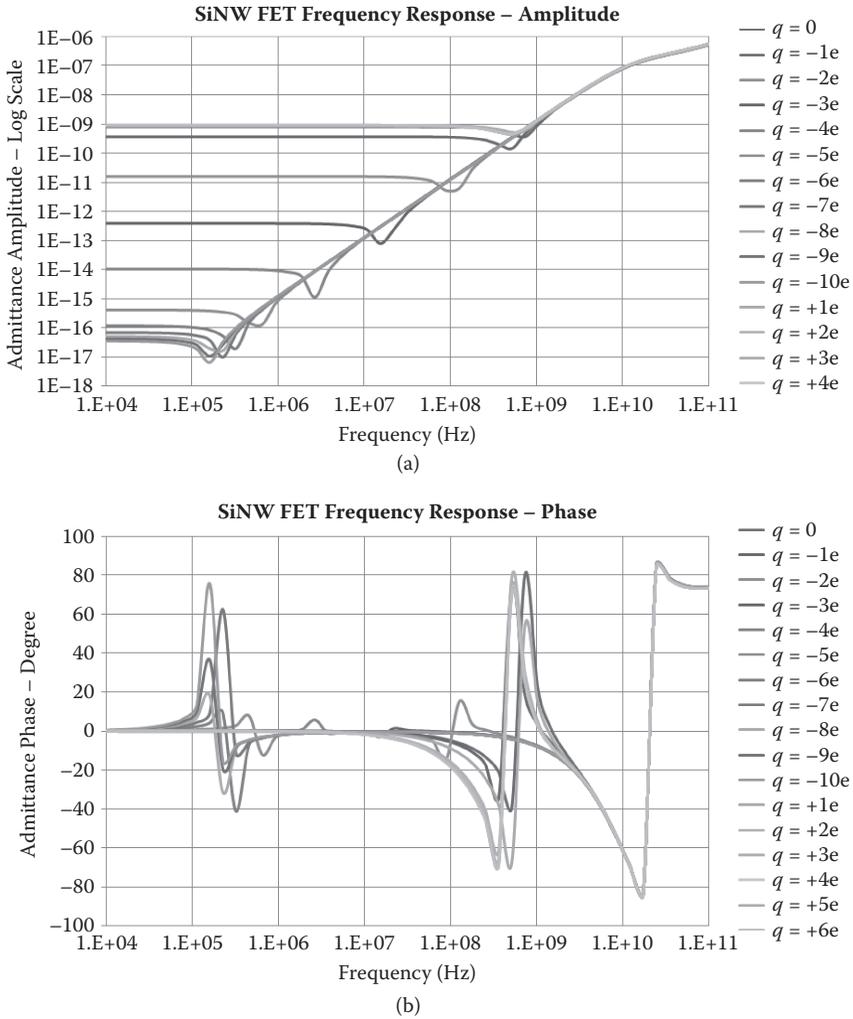

**FIGURE 11.5** SiNW FET frequency response as a function of gate charge. The nanowire has dimensions of 10 nm (W) × 10 nm (H). (a) Amplitude plot data of admittance. (b) Phase plot of admittance.

In [14], an AC signal analysis method was employed to detect antigen-antibody bonding with the subsequent determination of antigen concentration. From the frequency-based approach described, a signal was used to interrogate the device by sweeping from 10 kHz to many gigahertz while keeping the gate and source-drain voltage constant.

Figure 11.5 illustrates that both the amplitude (a) and phase (b) of the received signal versus frequency curves depend on the amount of charge attached to the nanowire. Furthermore, it can be illustrated that the corner frequencies in the amplitude plot correspond to those in the phase plot where the change in phase sign occurs. The actual corner frequency at zero charge can be determined once the nanowire device is fabricated. Then, when negative elementary charges attach to the nanowire surface, this will result in a significant difference in the nanowire's frequency response amplitude plot. We propose, therefore, that during the antigen detection process, by recognizing a shift in the corner frequency, the amount of negative charge on the nanowire surface can be determined once the nanowire is calibrated. The target antigen that is specifically bound by the antibody receptor on the modified surface, if detected, carries this charge.

To further study the frequency domain behaviors of silicon nanowire devices, AC signal analysis versus the nanowire device dimensions was performed. Sizes of 10 × 10 nm, 30 × 30 nm, and 50 × 50 nm (width by height) nanowires were compared with other parameters, such as length, doping profile, and aluminum pads, which were kept identical. Figure 11.6 illustrates the frequency response of three SiNW FET devices with different dimensions (i.e., width by height). It can be seen that an increase in channel width and height leads to an increase in device admittance, although the nanowire's high-pass filter behavior is retained. In addition, the change in the channel dimension also results in a change of the corner frequency of the transfer function. It can be seen from Figure 11.7 that while the charge varies from −1 $e$ to −10 $e$, as the cross-sectional area of a nanowire increases, we observe less distinction in the corner frequencies, making it more difficult to distinguish between the amounts of attached charge.

Sensitivity in these devices is ultimately dependent on how easy it is to recognize the difference in measuring signal during detection [18]. In traditional DC methods, detection is based on observing the change in conductance. This is difficult, especially when the absolute change is too small due to a low concentration of the target molecules, and requires sensitive low-noise amplifiers and high-resolution analog-to-digital converters, which is not ideal for low-cost and highly integrated systems. The frequency-dependent method of antigen detection, the task of discriminating the corner frequencies, seems to be closely located, as in Figure 11.6(c), and is straightforward as the difference is on the order of 100 MHz. As can be observed in Figure 11.7, this task becomes less complicated as the dimension of the nanowire channel shrinks. As a result, it can be concluded that there is an inversely proportional relationship between the device's sensitivity and its channel dimension.

Figure 11.7(a) also illustrates how the corner frequency varies as a function of attached charge and nanowire dimensions. Overall, with the same amount of charge, larger nanowire dimensions exhibit higher corner frequencies. It is also clear that at the lower end (i.e., smaller number of attached charges), corner frequencies linearly drop and diverge further as nanowire dimension gets smaller, making them significantly easier to be distinguished. However, as more surface charge is applied, the corner frequencies saturate. Moreover, the frequency saturation regions, which are the regions in Figure 11.6 where more charge does not result in corner frequency change, depend on nanowire channel size. Figure 11.7(b) also depicts a graph that shows the slope of the high-pass filter for an n-type nanowire device as described in the simulations. This graph demonstrates that regardless of the device width and height, these nanowire devices have approximately the same slopes.

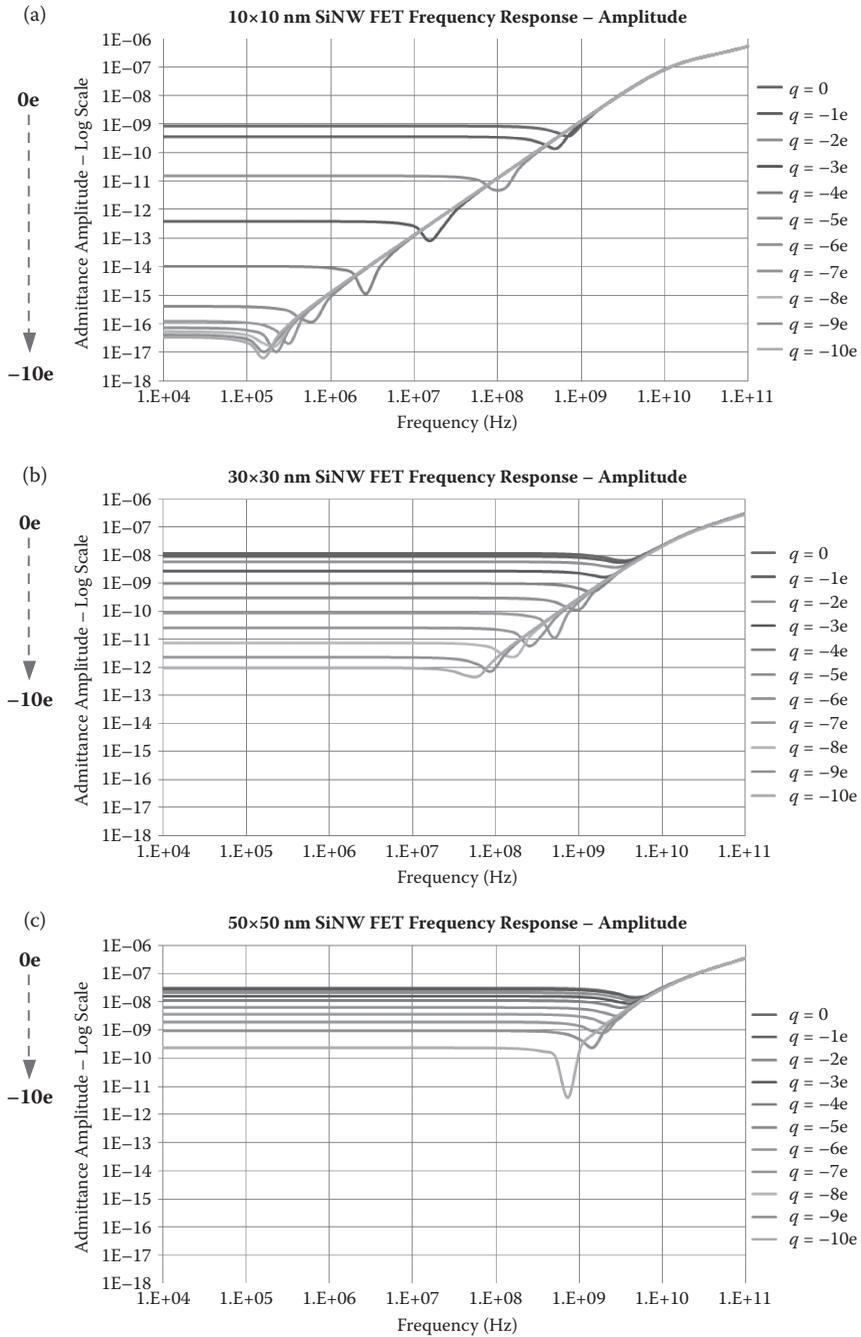

**FIGURE 11.6** SiNW FET frequency response as a function of gate charge and nanowire channel dimension: (a) $10 \times 10$ nm nanowire, (b) $30 \times 30$ nm nanowire, and (c) $50 \times 50$ nm nanowire. In these simulations, the nanowire length is 2.2 µm.

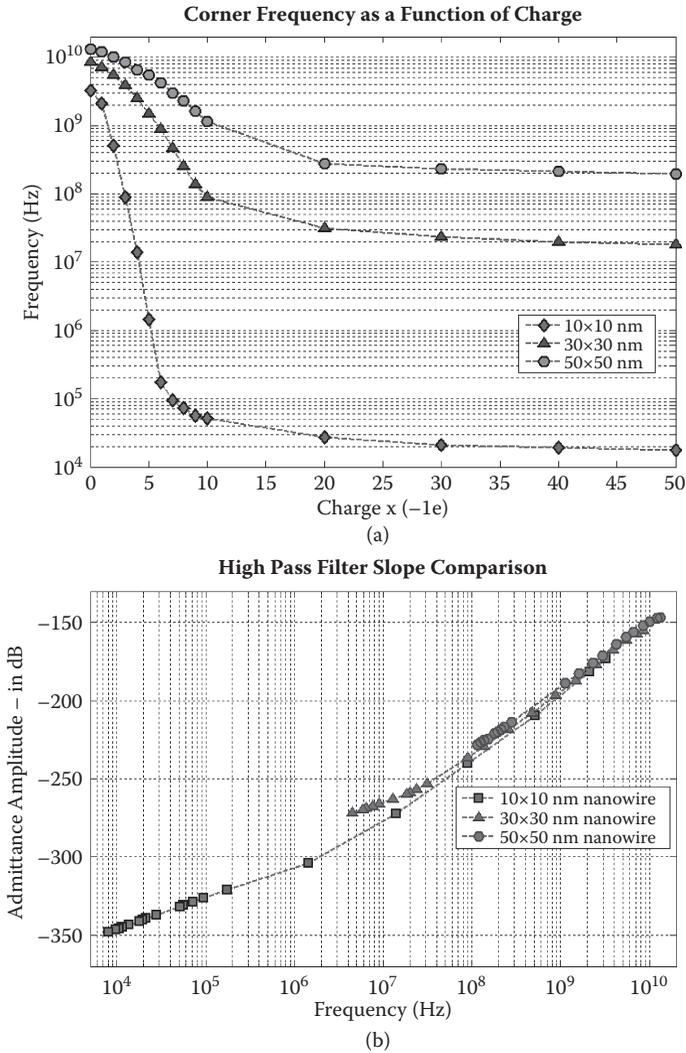

**FIGURE 11.7** (a) Corner frequencies as a function of charge attached and nanowire dimension showing how the corner frequency drops at low charge but saturates as the charge increases. (b) Comparison of nanowire devices' high-pass filter slope.

## 11.6 DETECTION CIRCUIT

The circuit proposed in Figure 11.8 can be used to determine the amount of charge that is attached to the nanowire. The circuit shown transmits an up-converted I/Q signal through the nanowire, which itself introduces a frequency-dependent phase change depending on the amount of charge attached to the nanowire. The receiving circuit amplifies this signal and down-converts it to a low frequency that is then sampled by two analog-to-digital converters.

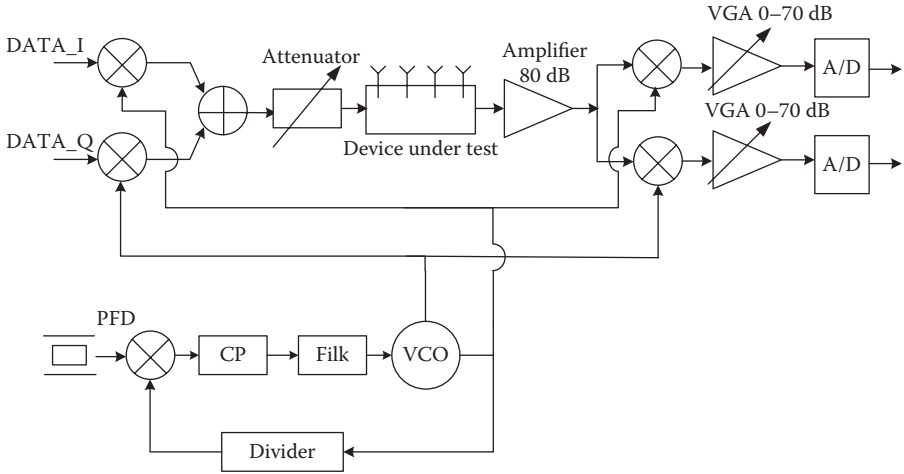

**FIGURE 11.8** A proposed detection circuit for the frequency-dependent method of biomolecule detection.

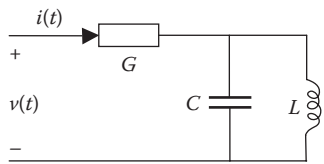

**FIGURE 11.9** Equivalent representation of the nanowire system.

This signal is then used to determine the phase change of the received signal relative to the transmitted signal. The charge is then determined by estimating the frequency at which there is a 45° phase change. This approach has the advantage of permitting robust identification of the attached charge when compared to methods trying to determine changes of conductance at DC, as it makes multiple measurements reducing the noise effect, but also avoids the need for accurate knowledge about the gain of the reception amplifiers. Here a ratio is used that is robust to low noise amplifier (LNA) amplification error. Calibration of the system is performed be applying a sample of known contribution. This would be done at the time of manufacture.

## 11.7 DETERMINATION OF CHANGE OF NANOWIRE PARAMETERS

The high-pass nature of the frequency responses revealed in Figures 11.5 and 11.6 suggests that the system can be represented by the circuit depicted in Figure 11.9. As a result, system identification techniques [19] can be utilized to identify the associated parameters (i.e., the conductance G, capacitance C, and inductance L) and, consequently, the corner frequency or phase change of the system. Together with the known frequency response properties obtained experimentally or through simulation, this result enables one to determine the amount of gate charge, leading to the detection of the antigen attached to the nanowire.

The system identification is based on measured current samples $\{i(n)\}$ in response to a certain excitation voltage sequence $\{v(n)\}$. According to Figure 11.8, the transfer function of the system in the $Z$-domain is given by

$$\frac{I(Z)}{V(Z)} = \frac{b_0 + b_2 Z^{-2}}{1 + a_1 Z^{-1} + a_2 Z^{-2}} \tag{11.1}$$

where
$b_0 = G$
$b_2 = G/(LC)$
$a_1 = G/C$
$a_2 = 1/(LC)$

The corresponding time domain expression of (11.1) is given by

$$i(n) = -a_1 i(n-1) - a_2 i(n-2) + b_0 v(n) + b_2 v(n-2)$$

For an $N$-sample input sequence $\{v(n), n = 0, 1, …, N - 1\}$, the current (output) samples $\{i(n), n = 2, 3, …, N - 1\}$ can be expressed in the following vector form:

$$y = Hb$$

where

$$y = \left[ i(2)\, i(3) … i(N-1) \right]^T$$

$$H = \begin{pmatrix} -i(1) & -i(0) & v(2) & v(0) \\ -1(2) & -i(1) & v(3) & v(1) \\ \vdots & \vdots & \vdots & \vdots \\ -i(N-2) & -i(N-3) & v(N-1) & v(N-2) \end{pmatrix}$$

$$b = \left[ a_1\, a_2\, b_0\, b_2 \right]^T$$

The least-squares [20] estimation of vector **b** is accomplished via

$$\hat{b} = H^+ y$$

where ( )+ denotes the Moore-Penrose pseudoinverse. Note that by setting the initial condition $i(0) = i(1) = 0$, both $H$ and **y** are known matrices. The estimation of vector **b** is accomplished via the following constraint least-squares minimization [20]:

$$\hat{b} = \arg\min_b \left\| H\mathbf{b} - y \right\|_2, \text{ subject to } b(4) = b(2) * b(3)$$

## 11.8 CONCLUSIONS

In this chapter, silicon nanowires that are compatible with CMOS fabrication processes have been described. It has been shown that these nanowires can be functionalized by conjugating monoclonal antibodies to their surface in order to build sensitive biochemical sensors. It has also been shown that by using frequency-based signals, all the necessary components to interrogate these nanowires can be built on low-cost CMOS processes.

## ACKNOWLEDGMENTS

The authors would like to acknowledge the generous support of National ICT Australia and the Centre for Neural Engineering at the University of Melbourne. This work was performed in part at the Melbourne Centre for Nanofabrication, an initiative partly funded by the Commonwealth of Australia and the Victorian Government.

## REFERENCES

1. Z. Gao, A. Agarwal, A.D. Trigg, et al. Silicon nanowire arrays for label-free detection of DNA. *Analytical Chemistry*, 79(9), 3291–3297, 2007.
2. C.-C. Wu, F.-H. Ko, Y.-S. Yang, et al. Label-free biosensing of a gene mutation using a silicon nanowire field-effect transistor. *Biosensors and Bioelectronics*, 25(4), 820–825, 2009.
3. X.T. Vu, R. Ghosh Moulick, J.F. Eschermann, et al. Fabrication and application of silicon nanowire transistor arrays for biomolecular detection. *Sensors and Actuators B: Chemical*, 144(2), 354–360, 2010.
4. C.-Y.H. Chih-Heng Lin, C.-H. Hung, Y.-R. Lo, C.-C. Lee, C.-J. Su, H.-C. Lin, F.-H. Ko, T.-Y. Huang, and Y.-S. Yang. Ultrasensitive detection of dopamine using a polysilicon nanowire field-effect transistor. *Chemical Communications*, 5749–5751, 2008.
5. G.-J. Zhang, Z.H.H. Luo, M.J. Huang, et al. Morpholino-functionalized silicon nanowire biosensor for sequence-specific label-free detection of DNA. *Biosensors and Bioelectronics*, 25(11), 2447–2453, 2010.
6. J.H. Chua, R.-E. Chee, A. Agarwal, et al. Label-free electrical detection of cardiac biomarker with complementary metal-oxide semiconductor-compatible silicon nanowire sensor arrays. *Analytical Chemistry*, 81(15), 6266–6271, 2009.
7. C.-Y. Hsiao, C.-H. Lin, C.-H. Hung, et al. Novel poly-silicon nanowire field effect transistor for biosensing application. *Biosensors and Bioelectronics*, 24(5), 1223–1229, 2009.
8. G.-J. Zhang, L. Zhang, M.J. Huang, et al. Silicon nanowire biosensor for highly sensitive and rapid detection of Dengue virus. *Sensors and Actuators B: Chemical*, 146(1), 138–144, 2010.
9. J.-I. Hahm and C.M. Lieber. Direct ultrasensitive electrical detection of DNA and DNA sequence variations using nanowire nanosensors. *Nano Letters*, 4(1), 51–54, 2003.
10. G.-J. Zhang, J.H. Chua, R.-E. Chee, et al. Label-free direct detection of MiRNAs with silicon nanowire biosensors. *Biosensors and Bioelectronics*, 24(8), 2504–2508, 2009.
11. Z. Gengfeng, F. Patolsky, C. Yi, et al. Multiplexed electrical detection of cancer markers with nanowire sensor arrays. *Nature Biotechnology*, 23(10), 1294–1301, 2005.
12. F. Patolsky, G. Zheng, O. Hayden, et al. Electrical detection of single viruses. *Proceedings of the National Academy of Sciences of the United States of America*, 101(39), 14017–14022, 2004.

13. J. Chang, A. Abidi, and C. Viswanathan. Flicker noise in CMOS transistors from sub-threshold to strong inversion at various temperatures. *IEEE Transactions on Electron Devices*, 41(11), 1965–1971.

14. T. Nguyen, W. Qiu, and E. Skafidas. Functionalised nanowire based antigen detection scheme using frequency based signals. *IEEE Transactions on Biomedical Engineering*, 59(1), 213–218, 2012.

15. C. Niemeyer (Ed.). Bioconjugation protocols. In *Methods in Molecular Biology*. Vol. 283. Humana Press, Clifton, NJ.

16. C. Yi, W. Qingqiao, P. Hongkun, et al. Nanowire nanosensors for highly sensitive and selective detection of biological and chemical species. *Science*, 293(5533), 1289, 2001.

17. E. Stern, J.F. Klemic, D.A. Routenberg, et al. Label-free immunodetection with CMOS-compatible semiconducting nanowires. *Nature*, 445(7127), 519–522, 2007.

18. P.R. Nair and M.A. Alam. Design considerations of silicon nanowire biosensors. *IEEE Transactions on Electron Devices*, 54(12), 3400–3408, 2007.

19. L. Ljung. *System Identification: Theory for the User*. Prentice Hall, Upper Saddle River, NJ, 1999.

20. S. Boyd and L. Vandenberghe. *Convex Optimization*. Cambridge University Press, Cambridge, 2004.